\documentstyle[12pt]{article}

\newcommand{\sect}[1]{\setcounter{equation}{0}\section{#1}}

\newcommand{\beq}{\begin{equation}}
\newcommand{\eeq}{\end{equation}}
\newcommand{\bea}{\begin{eqnarray}}
\newcommand{\eea}{\end{eqnarray}}

\renewcommand{\a}{\alpha}
\renewcommand{\b}{\beta}

\renewcommand{\d}{\delta}
\newcommand{\th}{\theta}

\newcommand{\pa}{\partial}
\newcommand{\g}{\gamma}
\newcommand{\G}{\Gamma}
\newcommand{\D}{\Delta}
\newcommand{\e}{\epsilon}

\renewcommand{\l}{\lambda}
\renewcommand{\L}{\Lambda}

\newcommand{\r}{\rho}

\newcommand{\s}{\sigma}
\renewcommand{\S}{\Sigma}
\renewcommand{\t}{\tau}

\renewcommand{\o}{\omega}
\renewcommand{\O}{\Omega}

\newcommand{\Db}{\bar{D}}

\newcommand{\Psib}{\bar{\Psi}}

\newcommand{\phib}{\bar{\phi}}

\newcommand{\ob}{\bar{\omega}}
\newcommand{\Phib}{\bar{\Phi}}

\newcommand{\chib}{\bar{\chi}}

\def\Mb{\kern 2pt\mathchoice
            {
             \vbox{\hrule width10pt height 0.4pt depth 0pt
                 \kern 1.2pt\hbox{\kern -2pt$\displaystyle M$}}}
            {
                 \vbox{\hrule width10pt height 0.4pt depth 0pt
                 \kern 1.2pt\hbox{\kern -2pt$\textstyle M$}}}
            {
\vbox{\hrule width6pt height 0.4pt depth 0pt
                 \kern 1.0pt\hbox{\kern -2pt$\scriptstyle M$}}}
            {
                 \vbox{\hrule width5pt height 0.4pt depth 0pt
                 \kern 0.8pt\hbox{\kern -2pt$\scriptscriptstyle M$}}}}

\def\Sb{\kern 2pt\mathchoice
            {
                 \vbox{\hrule width6pt height 0.4pt depth 0pt
                 \kern 1.2pt\hbox{\kern -2pt$\displaystyle S$}}}
            {
                 \vbox{\hrule width6pt height 0.4pt depth 0pt
                 \kern 1.2pt\hbox{\kern -2pt$\textstyle S$}}}
            {
                 \vbox{\hrule width3.5pt height 0.4pt depth 0pt
                 \kern 1.0pt\hbox{\kern -2pt$\scriptstyle S$}}}
            {
                 \vbox{\hrule width3pt height 0.4pt depth 0pt
                 \kern 0.8pt\hbox{\kern -2pt$\scriptscriptstyle S$}}}}

\def\Rb{\kern 2pt\mathchoice
            {
                 \vbox{\hrule width5.5pt height 0.4pt depth 0pt
                 \kern 1.2pt\hbox{\kern -2.5pt$\displaystyle R$}}}
            {
                 \vbox{\hrule width5.5pt height 0.4pt depth 0pt
                 \kern 1.2pt\hbox{\kern -2.5pt$\textstyle R$}}}
            {
                 \vbox{\hrule width3.5pt height 0.4pt depth 0pt
                 \kern 1.0pt\hbox{\kern -2.2pt$\scriptstyle R$}}}
            {
                 \vbox{\hrule width3pt height 0.4pt depth 0pt
                 \kern 0.8pt\hbox{\kern -2.2pt$\scriptscriptstyle R$}}}}

  \def\pp{{\mathchoice
          {
              \kern 1pt%
              \raise 1pt
              \vbox{\hrule width5pt height0.4pt depth0pt
                    \kern -2pt
                    \hbox{\kern 2.3pt
                          \vrule width0.4pt height6pt depth0pt
                          }
                    \kern -2pt
                    \hrule width5pt height0.4pt depth0pt}%
                    \kern 1pt
           }
            {
              \kern 1pt%
              \raise 1pt
              \vbox{\hrule width4.3pt height0.4pt depth0pt
                    \kern -1.8pt
                    \hbox{\kern 1.95pt
                          \vrule width0.4pt height5.4pt depth0pt
                          }
                    \kern -1.8pt
                    \hrule width4.3pt height0.4pt depth0pt}%
                    \kern 1pt
            }
            {
              \kern 0.5pt%
              \raise 1pt
              \vbox{\hrule width4.0pt height0.3pt depth0pt
                    \kern -1.9pt  
                    \hbox{\kern 1.85pt
                          \vrule width0.3pt height5.7pt depth0pt
                          }
                    \kern -1.9pt
                    \hrule width4.0pt height0.3pt depth0pt}%
                    \kern 0.5pt
            }
            {
              \kern 0.5pt%
              \raise 1pt
              \vbox{\hrule width3.6pt height0.3pt depth0pt
                    \kern -1.5pt
                    \hbox{\kern 1.65pt
                          \vrule width0.3pt height4.5pt depth0pt
                          }
                    \kern -1.5pt
                    \hrule width3.6pt height0.3pt depth0pt}%
                    \kern 0.5pt
            }
        }}

  \def\mm{{\mathchoice
                       {
                             \kern 1pt
               \raise 1pt    \vbox{\hrule width5pt height0.4pt depth0pt
                                  \kern 2pt
                                  \hrule width5pt height0.4pt depth0pt}
                             \kern 1pt}
                       {
                            \kern 1pt
               \raise 1pt \vbox{\hrule width4.3pt height0.4pt depth0pt
                                  \kern 1.8pt
                                  \hrule width4.3pt height0.4pt depth0pt}
                             \kern 1pt}
                       {
                            \kern 0.5pt
               \raise 1pt
                            \vbox{\hrule width4.0pt height0.3pt depth0pt
                                  \kern 1.9pt
                                  \hrule width4.0pt height0.3pt depth0pt}
                            \kern 1pt}
                       {
                           \kern 0.5pt
             \raise 1pt  \vbox{\hrule width3.6pt height0.3pt depth0pt
                                  \kern 1.5pt
                                  \hrule width3.6pt height0.3pt depth0pt}
                           \kern 0.5pt}
                       }}

\def\pd{{\kern0.5pt
                   + \kern-5.05pt \raise5.8pt\hbox{$\textstyle.$}\kern
0.5pt}}

\def\pmd{{\kern0.5pt
                  \pm \kern-5.05pt \raise6.3pt\hbox{$\textstyle.$}\kern1.5pt}}

\def\md{{\mathchoice
   {
      {{\kern 1pt - \kern-6.2pt \raise5pt\hbox{$\textstyle.$}\kern 1pt}}}
    {
      {{\kern 1pt - \kern-6.2pt \raise5pt\hbox{$\textstyle.$}\kern 1pt}}}
    {
      {\kern0.5pt - \kern-5.05pt \raise3.4pt\hbox{$\textstyle.$}\kern0.5pt}}
    {
      {\kern0.5pt - \kern-5.05pt \raise3.4pt\hbox{$\textstyle.$}\kern0.5pt}}}}

\newcommand{\ad}{{\dot{\alpha}}}
\newcommand{\bd}{{\dot{\beta}}}
\newcommand{\Del}{\nabla}
\newcommand{\Delb}{\bar{\nabla}}

\begin{document}

\topmargin 0pt
\oddsidemargin 5mm

\renewcommand{\Im}{{\rm Im}\,}
\newcommand{\NP}[1]{Nucl.\ Phys.\ {\bf #1}}
\newcommand{\PL}[1]{Phys.\ Lett.\ {\bf #1}}
\newcommand{\NC}[1]{Nuovo Cimento {\bf #1}}
\newcommand{\CMP}[1]{Comm.\ Math.\ Phys.\ {\bf #1}}
\newcommand{\PR}[1]{Phys.\ Rev.\ {\bf #1}}
\newcommand{\PRL}[1]{Phys.\ Rev.\ Lett.\ {\bf #1}}
\newcommand{\MPL}[1]{Mod.\ Phys.\ Lett.\ {\bf #1}}
\renewcommand{\thefootnote}{\fnsymbol{footnote}}
\newpage
\begin{titlepage}

\begin{flushright}
IFUM--539--FT \\
BRX--TH--389\\
hep-th/9610001\\
September 1996
\end{flushright}

\vspace{2cm}
\begin{center}
{\bf{\large  QUANTIZING $N=2$ MATTER--SUPERGRAVITY SYSTEMS}} \\
\vspace{2cm}
{\large M.T. Grisaru}\\
\vspace{.2cm}
{\em Physics Department, Brandeis University, Waltham, MA 02254, USA}\\
\vspace{.5cm}
and\\
\vspace{.5cm}
{\large  A. Santambrogio and D. Zanon} \\
\vspace{.2cm}
{\em Dipartimento di Fisica dell' Universit\`{a} di Milano and} \\
{\em INFN, Sezione di Milano, Via Celoria 16, I-20133 Milano, Italy}\\
\end{center}
\vspace{2cm}
\centerline{{\bf{Abstract}}}
\vspace{.5cm}

We consider $N=2$ supergravity coupled to $N=2$ Yang--Mills matter
and discuss the nature  of one--loop  divergences.
Using   $N=1$ superfields and superspace methods, we describe the
quantization of the system in the abelian case.

\vfill
\end{titlepage}
\renewcommand{\thefootnote}{\arabic{footnote}}
\setcounter{footnote}{0}
\newpage

\sect{Introduction}
Recently there has been considerable  activity in
studying matter--supergravity lagrangians as effective low--energy
representations of superstring theories. This has been done both in
component and in $N=1$ superspace formulations, but primarily at the
classical level. Of special interest have been theories with $N=2$
supersymmetry,  both from a phenomenological point of view \cite{pheno} and
because of the interesting geometrical structures which emerge there,  with
particular emphasis on special and quaternionic geometry
\cite{torino}.
It is expected that below a certain energy
scale these theories can be used without invoking the full string
technology.  However, they necessarily contain
nonrenormalizable interactions. Thus in order to have realistic and
phenomenologically interesting models one has to show that naturalness
is not spoiled and the hierarchy problem is
not affected by radiative corrections.  In particular it is  important to
study the  ultraviolet cutoff dependence for  these effective theories
derived from strings.

At the quantum level, it is well-known that, whereas pure supergravity is
on--shell one-- and two--loop finite \cite{GVN}, in  (super)gravity--matter
systems divergences appear already at the one--loop level \cite{veltman}.
Thus, if these effective
theories are to represent
superstring models to  moderately high scales, it is of interest to know the
nature
of these divergences. {\em A priori}, one expects to encounter
in the  one--loop effective actions for matter fields  quadratically
divergent  quantum supergravity
corrections, and indeed, recent component calculations  in $N=1$ models have
verified their existence \cite{gaillard}. Other studies \cite{bagger},
have indicated that the presence of quadratic divergences, even in a ``hidden''
sector, can have a destabilizing effect  on the
hierarchy problem.

Our interest in these issues stems from the expectation that $N=2$
matter systems coupled to $N=2$ supergravity may display better
high--energy behavior.
Since some of the favored
string--inspired models have $N=2$ supersymmetry before breakdown,
at a lower scale, to $N=1$ and ultimately to $N=0$, this problem
is of more than academic interest.

We consider $N=2$ supergravity coupled to $N=2$ Yang--Mills matter
\cite{rocek,laba}.
First we  present a general superspace power counting
argument, on the basis of $N=2$ supersymmetry,
to determine the nature of  the one-loop divergences.
Then we address the problem of quantizing the system. We have to deal with
several gauge
fields, and their quantization  consists primarily in  fixing gauges and
determining the corresponding ghost actions. We work in $N=1$ superspace
because at the present time no suitable $N=2$ description of the supergravity
sector is available.  In this formalism, the gauge--fixing violates $N=2$
invariance. Consequently, to
extract properties which follow from the full  supersymmetry we must look at
on--shell quantities which do not depend on the gauge parameters or the form
of the gauge--fixing functions. We  fix gauges in the simplest
and most convenient way, avoiding as much as possible the introduction of
couplings of the ghost fields to the physical fields.

Our paper is organized as follows: in section 2, we describe the
general  power counting argument
regarding the nature of one--loop divergences.
In section 3, we present our
$N=2$ model consisting of an abelian vector multiplet coupled to
supergravity. For this system a complete $N=1$ superspace
description has been given by Labastida {\em et al} \cite{laba}.
It contains an abelian $N=1$ vector multiplet and a scalar multiplet,
coupled to nonminimal $n=-1$ supergravity, as well as to a gravitino
multiplet. In order to construct local matter--supergravity
interactions, the scalar multiplet has to be described by a complex scalar
prepotential rather than the customary chiral scalar superfield.

The quantization of the matter multiplets, i.e. the vector multiplet
and the above mentioned complex scalar is straightforward and briefly
discussed in section 4.  In section 5 we study the
quantization of the $N=2$ supergravity system.
We describe the gravitino multiplet by means of a
complex spinor superfield, and  chiral scalar and real scalar
compensating superfields. We fix the various gauge invariances
by appropriate choices of gauge--fixing functions, which
allow  putting the quadratic action in standard form and obtaining simple
propagators.
Finally we discuss the quantization of
 nonminimal $n=-1$ supergravity, described by a real
vector superfield and a complex linear superfield compensator.
In general the gauge--fixing procedure  requires introducing
corresponding Faddeev--Popov ghosts. We have made all our
gauge choices flat with respect to both the Yang--Mills and the
supergravity fields, so that we need not worry about coupling of the
Faddeev--Popov ghosts to physical multiplets except in the case of
nonlinear variations of the fields. For the system under consideration
this happens only for the $n=-1$ supergravity superfields. We discuss
in detail the quantization of the corresponding Faddeev--Popov
lagrangian at the end of section 5.

One feature of the quantization process is the introduction of
``catalyst'' fields \cite{book}, a  procedure for recasting kinetic
lagrangians,
gauged--fixed but not in convenient form,  to a form which leads to standard
propagators. Unlike most of the ghost fields mentioned above, many of the
catalysts end up  being coupled to the physical fields.
Also, it is well--known that the
quantization of the linear superfield introduces an infinite tower
of ghosts. However, since the latter do not couple to the physical
fields, we can avoid discussing the difficulties introduced
by them.
In the Appendix we have listed some useful formulas.

We use superspace notations and conventions as in ref. \cite{book}.

\sect{On quadratic divergences for $N=2$ Yang--Mills in presence
of $N=2$ supergravity}

By ordinary power counting  -- the gravitational coupling constant $\kappa$ has
dimensions of (mass)$^{-1}$ --  conventional lagrangian models of
(super)gravity--matter systems are not renormalizable. The effective action for
such systems is on--shell finite at the one--loop level
when restricted to the case of  external gravitational lines only
\cite{veltman},
and at the two--loop level as well for the corresponding
supergravity case \cite{GVN}.
However, divergences appear as soon as one
considers external matter lines, even on shell. At best, one may hope
that the degree of divergence is only logarithmic, so that such models
can be used as
low--energy effective actions from strings even at moderate
energies.

As mentioned in the introduction, it has been shown that  supergravity--matter
systems with $N=1$ supersymmetry develop quadratic divergences already at the
one--loop level \cite{gaillard},
and these can have deleterious effects on the validity of
string--inspired lagrangian models beyond energies where classical
considerations are sufficient.  Specifically, it has been shown that the
effective action for scalar multiplets or vector multiplets, with radiative
corrections due to the exchange of supergravity fields,  depends quadratically
on the ultraviolet cutoff  and this can destroy  properties of
supersymmetric systems such as naturalness and the solution to the hierarchy
problem, even when they occur in a ``hidden'' sector  \cite{bagger}.
We present here a  power counting argument, based on $N=2$ supersymmetry
and gauge invariance,
that indicates that the situation may be  better  in the case of  the one--loop
effective action for $N=2$ Yang-Mills  with radiative corrections  due to $N=2$
supergravity fields. Whereas terms involving $N=2$ chiral superspace
integrands of the form ${\cal F}(W)$ -- here $W$ is the $N=2$ Yang-Mills
field strength --are quadratically divergent, full superspace terms
are at most logarithmically divergent.  In particular, for the chiral multiplet
component of the $N=2$ Yang-Mills superfield, one might
encounter quadratic divergences  only for terms of the form
$\bar{\omega} {\cal F}(\omega ) + h.c.$.  For the model we are
considering in this paper, only the two--point function
 $\bar{\omega} \omega$  would be  quadratically divergent.

Our power counting arguments assume that the amplitudes under consideration do
have manifest  {\em global} $N=2$ supersymmetry. In practice, however, since a
suitable  $N=2$ superspace formulation of supergravity is not
available, explicit calculations have to be performed in an  $N=1$ (or
component) formalism.
In general, any calculation which is not manifestly $N=2$ supersymmetric (i.e.
not done in terms of $N=2$ superfields) will involve breaking of the
(extended)
supersymmetry, either because auxiliary fields have been eliminated, or because
the fixing of the local gauge invariances has to be done in a manner
which does not respect it.   Specifically for the case under consideration, as
we will discuss below, the fixing of the various gauge invariances has to be
done separately for the $N=1$ members of the $N=2$ multiplets and this leads
to some explicit breaking of the $N=2$  {\em global} invariance. Therefore,
in practice we have to restrict ourselves to the
computation of on--shell, gauge invariant and gauge indepedent quantities, i.e.
S--matrix elements.

For amplitudes involving external Yang--Mills fields,
$N=2$ supersymmetry and gauge invariance imply that the divergent part  has to
be a local expression depending on the Yang--Mills field strength, the $N=2$
chiral superfield $W$, in an integral which is either over chiral superspace
with a $d^4 \th$
measure, or over full superspace, with a $d^8 \th \equiv d^4 \th d^4
\bar{\th}$ measure.
We are using spinor coordinates $\th^\a_1$, $\th^\a_2$ and their complex
conjugates,
with $\Db^\ad_i W=0$.
 Besides the fields and integration measures (including the space-time measure
$d^4x$), other dimensionful quantities are the gravitational coupling $\kappa$,
and an ultraviolet cutoff $\Lambda$. The (mass) dimensions of the various
quantities are listed below:
\bea
\left[ d^4x \right] &=& -4 ~~~~~~~~~~~~~~ \left[ d^4 \th \right]  = 2
\nonumber\\
\left[ W\right] &=& ~1 ~~~~~~~~~~~~~~~~~~~ \left[ \kappa \right] = -1
\nonumber\\
\left[ \Lambda \right] &=&~ 1
\eea
Each (super)gravitational  internal propagator brings with it a power of
$\kappa^2$ and it is easy to see that, in the absence of any matter
self--interactions, a one--loop diagram with $2n$ external
$W$--lines involves $n$ supergravity propagators. Thus, allowed dimensionless
local expressions are of the form
\beq
\int d^4x d^4 \th ~W^{2n} ~ \kappa^{2n} \Lambda^2
\label{four}
\eeq
involving chiral (or antichiral) integrals, and
\beq
\int d^4x d^8 \th ~ W^n \bar{W}^n ~\kappa^{2n}  \ln \Lambda
\label{eight}
\eeq
with full superspace integrals,
or expressions with (spinor or space-time) derivatives, which are more
convergent

Upon reduction to $N=1$ superspace, these expressions can be rewritten in term
of the superfields of the $N=1$ vector and scalar multiplets, $V$, $\o$,  which
make up the $N=2$ Yang--Mills multiplet. We are particularly interested in
the contributions to the scalar multiplet effective action. The results of this
reduction are well known:
one defines $N=1$ components by
\beq
\o = W| ~~~,~~~~W^\a = - D_2^\a W|
\eeq
where the bar indicates evaluation at $\th^{\a}_2=\bar{\th}^{\ad}_2=0$ and
rewrites the
integration measure in terms of $N=1$ measures, with the replacement $d^2 \th_2
\rightarrow
\frac{1}{2} D^\a_2 D_{2\a}$, etc.
In particular, as far as the pure scalar multiplet is concerned, the chiral
superspace integral in (\ref{four})  leads to expressions of the form
$ 2n \bar{\o} \o^{2n-1}$.
The full superspace expression in (\ref{eight}) on the other hand, leads to
chiral superfield contributions of the form (in the abelian case, for the
nonabelian case see also ref. \cite{bernard})
\beq
 n^2\bar{\o}^{n-1} \o^{n-1} \bar{D}^2 \bar{\o} D^2 \o +
n\bar{\o}^{n-1} \o^n  \Box \bar{\o} +\frac{n}{2}(n-1) \o^n
\bar{\o}^{n-2} \pa^a \bar{\o} \pa_a\bar{\o}
\eeq
For theories  such as we will consider below  in which the Green's
functions have equal number of external $\o$ and $\bar{\o}$ fields,
quadratic divergences are possibly present only in the chiral superfield
two--point function. The higher--point terms are
at most logarithmically divergent and
contain derivatives of the chiral superfields.

These conclusions are applicable whenever one can argue that the results of a
calculation are gauge invariant and $N=2$ supersymmetric.  In our context, this
means on--shell,
gauge invariant and gauge independent quantities.  Thus, in an $N=1$
calculation they
apply to  scalar multiplet  (and by $N=2$ supersymmetry vector multiplet)
scattering amplitudes
which should have at most logarithmic divergences.

\sect{$N=1$ superfield description of the $N=2$ vector multiplet
coupled to $N=2$ supergravity}

$N=2$ extended supergravity in $N=1$ superspace
is described by  the  nonminimal,  $n=-1$, version of $N=1$ supergravity,
and a gravitino multiplet. The former consists of  the vector
superfield  $H_{\a \ad}$
and a  complex linear compensator $\Upsilon$. For the latter,
a convenient
description is by means of a  spinor superfield $\phi_\a$
and  scalar compensators $\cal{V}$ (real),
and $\Phi$ (chiral). The full $N=2$ action has been constructed
in ref. \cite{rocek}.

The $N=2$ abelian vector multiplet consists
of an $N=1$ vector multiplet, described
by a scalar  (prepotential) superfield  $\O$, and a scalar multiplet
described
by a chiral superfield $\o$. However, in order to couple this matter
system to $N=2$ supergravity it is necessary to
solve the chirality constraint in terms of a (gauge) prepotential $\Psi$,
$$\o = \bar{\Del}^2 \Psi$$
The complete  $N=1$ action for the coupled system, as given in
refs. \cite{rocek}, \cite{laba} is
\bea
 S&=& -\frac{1}{2 \kappa^2} \int d^4x d^4\theta~ E^{-1}\left\{
2C+ \left[ \phi^{\a}( \l_{\a}+{\cal{W}}_{\a})C \right. \right. \nonumber\\
&&~~~~~~~~~~~\left. \left.
+(\bar{x}+x\bar{B})(\phi^{\a} T_{\a}
+\bar{\Phi}) +\frac{i}{2} {\cal{W}}^{\a} \Del_{\a}
{\cal{V}} + h.c. \right] \right\} \nonumber\\
&&+\int d^4x d^4\th~ E^{-1} \left\{ \o \bar{\o}[C+ \frac14(x \bar{v}
 +\bar{x}v)] \right. \nonumber\\
&&~~~~~~+\left.  \frac14 \left[ \G^\a (N_\a^{~\b} \o_\b +C \l_\a \o
+\bar{v}(\Del_\a +T_\a)\o ) +
\frac12 \o \bar{v} \Del^\a \G_\a +h.c.\right] \right\}
\label{action}
\eea
where $\Del_A$, with  $A= \a, \ad , a \equiv(\a \ad)$, denote  suitably defined
$N=1$ supergravity covariant derivatives with connections including
some gravitino multiplet contributions. ${\cal{W}}_\a\equiv
i\bar{\Del}^2 \Del_\a {\cal{V}}$ is the ${\cal V}$ field--strength,
$E$ is the $N=1$ supergravity vielbein determinant,
$\G_\a$ and $\omega_{\a}$ are
supercovariantized
vector multiplet $U(1)$ spinor  connection and field strength, respectively,
 and the remaining quantities,
$B$, $C$, $v$ and $x$,
and $N_\a^{~\b}$, $\l_\a$ and $T_\a$ are composed of fields in the
gravitino multiplet and their derivatives.
We give a summary  of some  relevant quantities:
\bea
\G_\a &=& \Del_\a \O - 2 \psi_{\a \b} \Del^\b \bar{\Psi} \nonumber\\
\psi_\a^{~\b}&=&\Del_\a \phi^\b +\d_\a^{~\b} \bar{\Phi} \nonumber\\
x &=& \psi_{\a}^{~\a}
\nonumber\\
B&=&\frac12 \psi^{\a\b} \psi_{\a\b} \nonumber\\
 v&=&x +\bar{x}B \nonumber\\
 C&=&1 -B\bar{B} \nonumber\\
N_\a^{~\b}&=&C \d_\a^{~\b} +\bar{v} \psi_\a^{~\b} \nonumber\\
\l_\a &=& -\bar{\Del}_\ad \Del_\a \bar{\phi}^\ad+
\bar{\Del}^2\phi_\a +\Del_\a \Phi + {\cal{W}}_\a
+higher ~order~ terms \nonumber\\
\o_\a&=&\frac{i}{4} \Delb ^\ad \G_{\a \ad} - \frac{i}{4} \Del_{\a \ad}
\bar{\G}^\ad - \l_\a \o
-\frac12 [\psi_{\ad}^{~\ad} \Del_\a -
\Del_\a \psi_{\ad}^{~\ad}]\o
\nonumber\\
&&~~~~~~~~~~~~~~~~~~~~~~~~~~~~+~ higher~ order~ terms
[6~\label{definitions}
\eea
The complete definitions can be found in  ref. \cite{laba}.

In addition to the $N=1$ supersymmetry invariance which is implicit in our use
of $N=1$ superfields, the action in (\ref{action}) is invariant under
a second supersymmetry transformation.
With  spinor parameter $\e_\a$, it acts on the matter fields as
\bea
\d \o &=& - \e^\a \o_\a \nonumber\\
\d \o_\a &=& \e_\a \Sigma - i\bar{\e}^\ad [\Del_{\a \ad} \o
+\psi_{\a \ad}^{~~~\b} \o_\b ]
\nonumber\\
\d \G_\a &=& 2\bar{\o} \e_\a +\bar{\e}^\ad[ \frac12
\bar{\lambda}_\ad \G_\a +i \psi_\a^{~\g}\G_{\g \ad}]
\eea
with $\G_{\a \ad}$ the $U(1)$  Yang-Mills vector connection,
and    $\psi_{\a \ad}^{~~~\b}$ expressible
in terms of derivatives of the gravitino fields.  $\Sigma$ is
a function of the fields in the matter and gravitino
multiplets. Their expressions to lowest order in the supergravity
fields are given by
\bea
i\G_{\a\ad}&=& \Del_\a \bar{\G}_\ad +\Delb_\ad \G_\a+\dots \nonumber\\
i\psi_{\a\ad}^{~~~\b}&=&- \frac12 \d_\a^{~\b} \bar{\l}_\ad + \Delb_\ad
\psi_\a^{~\b}+\dots\nonumber\\
\S&=&
2\Delb^2\bar{\o} -\l^\a \o_\a + \Delb^\ad \psi_\ad^{~\bd} \bar{\o}_\bd
-\psi^{(\ad \bd)} \Delb_\ad \bar{\o}_\bd+\dots
\eea
The second supersymmetry transformation laws for the supergravity
fields can be found in ref. \cite{rocek}.

To quantize the theory, since we have gauge fields, $\O$ and $\Psi$
as matter prepotentials, $H_a$, $\phi_\a$ as fundamental supergravity
fields, we need fix the various gauge invariances. In principle one
would like to perform the quantization in a way that maintains
explicit $N=2$ supersymmetry. This could be achieved  by
introducing a quantum--background splitting for the various superfields
so that gauges could be fixed in a $N=2$ background covariant way.
In practice the analysis of the quantum and background separate invariances
becomes so cumbersome that this approach is not easy to  implement.
Thus we apply a low--brow procedure  by separately gauge--fixing
the various $N=1$  prepotentials. This is sufficient
for calculating S--matrix elements. These on--shell quantities
are the relevant ones at the quantum level: they are uniquely defined, being
independent of the definite gauge choices made in the course of the
quantization. Our starting point is the quadratic part of
the action in (\ref{action})
\beq
S^{(2)}\equiv S^{(2)}_m +S^{(2)}_G +S^{(2)}_g
\label{quadraticaction}
\eeq
where the subscripts $m$, $G$ and $g$ indicate matter, gravity and
gravitino fields respectively. We present the details of the
gauge--fixing procedure in the next two sections.

\sect{Quantization of $N=2$ Yang--Mills}

We concentrate here on the matter part of the action. As already mentioned,
 the basic superfields are two
unconstrained complex prepotentials $\O$ and $\Psi$ which
appear in the action (\ref{action}) only through the quantities
\beq
\o=\bar{\Del}^2 \Psi \qquad \qquad \qquad
\G_\a = \Del_\a \O - 2 \psi_{\a \b} \Del^\b \bar{\Psi}
\label{YMprep}
\eeq
In addition to the usual gauge invariance of the vector multiplet
\beq
\d\O=\bar{\L}   \qquad \qquad \qquad \Del_\a \bar{\L}=0
\label{chiral}
\eeq
and the $U(1)$ invariance
\beq
\d\O=\kappa \qquad \qquad \qquad \qquad \d\Psi=0
\label{abelian}
\eeq
with $\kappa$ real,
the definitions in (\ref{YMprep}) exhibit the  extra invariance under
the following transformations with spinor parameter $\chi_\a$ \cite{laba}
\beq
\d\Psi=\Delb^\ad \chib_\ad \qquad \qquad \qquad
\d\O=2\phi_\b \Del^\b \Del^\g \chi_\g -2\Phib \Del^\g
\chi_\g
\label{extra}
\eeq
The $U(1)$ $\kappa$--transformation can be used to reach a
gauge in which $\O$ is purely imaginary; once this has been achieved
every transformation that would take one out of this gauge has to
be followed by a compensating $\kappa$--transformation, with $\kappa$
chosen so as to cancel the real part of $\O$. In the following
we can then set $\O=-\bar{\O}\equiv iV$.

In all situations in which  $\O$ ( or $V$ ) and $\o=\bar{\Del}^2\Psi$
propagate in the loops, but no  bare quantum  $\Psi$ is present,
there is no need to solve the chirality
constraint for the quantum $\o$. In this case
one only has to gauge fix the standard invariance in (\ref{chiral}).
This can be done,
while maintaining the extra invariance in (\ref{extra}), by choosing
a gauge--fixing term of the form
\beq
S^m_{GF}= -\frac{1}{4} \int  d^4x d^4\theta~ \Db^\ad \bar{\G}_\ad
D^\a \G_\a
\label{YMGF}
\eeq
and combining this term with the corresponding classical
part of the action
\beq
 S_m= \frac{1}{16} \int d^4x d^4\th~ \G^\a[ \Db^\ad(D_\a
\bar{\G}_\ad +\Db_\ad \G_\a) -i\pa_{\a \ad}\bar{\G}^\ad ]
\label{YMclass}
\eeq
We note that substituting in (\ref{YMclass}) the explicit  forms of
$\G^\a$, $\bar{\G}^\ad$, the gauge field $V$ appears with
the conventional  lagrangian $\frac12 V D^\a \Db^2 D_\a V$.
In the same manner (\ref{YMGF}) contains the standard gauge-fixing
function $-D^2V \Db^2V$ which converts it into the Feynman gauge
form $-\frac12 V \Box V$.

If we are in a situation in which the field $\Psi$ itself has to be
treated as a quantum field, then necessarily we have to break the
symmetry in (\ref{extra}). This is achieved by choosing as gauge--fixing
function
\beq
F_\a =D_\a \Psi
\label{psi}
\eeq
and introducing a gauge--fixing term of the form
\beq
-\int d^4x d^4\th~ \bar{F}^\ad(\frac14 \Db_\ad D_\a-i \pa_{\a\ad}) F^\a
=\int d^4x d^4\th~\bar{\Psi}(\Db^2 D^2 - \Db^\ad D^2 \Db_\ad)\Psi
\label{GFPsi}
\eeq
Again, when combined with the classical term $\ob \o= D^2\Psib \Db^2\Psi$,
it gives rise to a standard $\Psib \Box \Psi$ kinetic term.

Faddeev--Popov ghosts should be introduced corresponding to the gauge--fixing
in (\ref{YMGF}), but since $V$ is abelian they do not couple to the physical
fields and therefore they are irrelevant. The same conclusion can
be reached for the ghosts of the $\Psi$ superfield. In this case
the gauge variation of $\Psi$ has zero --modes,  $\d \bar{\chi}_{\ad} =
\Db^{\bd}
\bar{\chi}_{(\ad \bd )}$  (see (\ref{extra}))
and  the gauge--fixing of the ghost lagrangian will introduce
an infinite tower of ghosts. The same situation occurs in the quantization
of the supergravity multiplet, to be discussed below. However,
with the flat gauge fixing in (\ref{psi}), they are completely decoupled and
do not contribute to physical amplitudes.

\sect{Quantization of  $N=2$ supergravity}

\subsection{Quantization of the gravitino multiplet}

We begin by studying the gauge--fixing procedure for the
$N=1$  gravitino multiplet, described by the general
spinor superfield $\phi_\a$,
and a pair of compensators that allow us to write a local action
(see ref. \cite{book}, sec. 4.5.e; for our purpose, the most
convenient choice of compensators
uses  a real
vector ${\cal{V}}$ and a chiral scalar $\Phi$). The corresponding
quadratic part of the action contained in (\ref{action}),
( also {\em cf.} \cite{book},
eq. (4.5.36)), can be written in the form
\bea
S_g^{(2)} &=& -\frac12 \int d^4x d^4\th \left[ \phi^\a
( \Db^\ad D_\a \phib_\ad +\Db^2 \phi_\a
+2{\cal{W}}_\a +D_\a \Phi )\right. \nonumber\\
&~&\left. ~~~~~~~~~~~~~~~~~~~~~~~+ \phib^\ad ( D^\a \Db_\ad \phi_\a +D^2
\phib_\ad
+2\bar{\cal{W}}_\ad +\Db_\ad \Phib )  \right. \\
&~&\left.~~~~~+ (D_\a \phi^\a +2 \Phib )\Phi +(\Db_\ad \phib^\ad +2\Phi )
\Phib +\frac{i}{2}
(D^\a {\cal{V}} {\cal{W}}_\a -\Db^\ad {\cal{V}} \bar{\cal{W}}_\ad ) \right]
\nonumber
\eea
with ${\cal{W}}_\a \equiv i\Db^2D_\a {\cal{V}}$. It is invariant under
the gauge transformations
\bea
\d \phi_\a &=& \L_\a +D_\a Z~~~~~~~~~~~~~~~~~~~~~~~~~~~~~~~~~\Db_\ad \L_\a=0
\nonumber\\
\d \cal{V}&=& i(Z - \bar{Z}) +\L+\bar{\L}~~~~~~~~~~~~~~~~~~~~~~~
\Db_\ad \L=0  \nonumber\\
\d \Phi &=& \Db^2 \bar{Z}
\label{grav}
\eea
where $Z$ is a general superfield.

We use the following gauge-fixing functions:
\bea
F_\a&=&\Db^2 \phi_\a +a~ { \cal{W}}_\a \nonumber\\
F&=&D^\a\phi_\a + b~\bar{\Phi} \nonumber\\
G &=& D^2 \cal{V}
\eea
and the gauge-fixing lagrangian
\beq
L_{GF} = c~ F^\a \frac{D^2}{\Box}F_{\a}
+c~\bar{F}^\ad\frac{\Db^2}{\Box}\bar{F}_{\ad}+d~\bar{F}F+e~\bar{G}G
\eeq
(Nonlocal gauge-fixing terms are a common feature of the quantization procedure
in superspace.) After some algebra, we find that with the choice
\beq
a=d=1\nonumber \qquad \qquad \qquad b=e= -1 \qquad
\qquad \qquad c=\frac12
\eeq
the lagrangian takes the form
\beq
L_g^{(2)}+L_{GF} = i\phi^\a \pa_{\a \ad}\phib^\ad -\Phib \Phi -
\frac12{\cal{V}}\Box {\cal{V}}
\eeq
Thus we obtain standard propagators
\bea
\langle \phi_\a (x, \th ) \phib _\ad (x', \th ')  \rangle
&=& i \frac{\pa_{\a \ad}}{\Box}
 \d^4(x-x') \d^4(\th - \th ')  \nonumber\\
\langle \Phi (x, \th ) \Phib  (x', \th ')  \rangle
&=&  \frac{1}{\Box}
 \d^4(x-x') \d^4(\th - \th ')  \nonumber\\
\langle {\cal{V}}(x, \th ) {\cal{V}}(x', \th ')  \rangle &=& \frac{ 1}{\Box}
 \d^4(x-x') \d^4(\th - \th ')
\eea

Since we have considered flat (with respect to Yang--Mills
and supergravity) gauge-fixing functions and  the gauge
transformations in (\ref{grav}) are all linear, the ghost
fields are not interacting and can be dropped.

\subsection{Quantization of $N=1$  nonminimal supergravity}

We discuss now gauge-fixing for the $N=1$ supergravity sector.
Nonminimal $n=-1$ supergravity is described by the real vector
superfield $H^a$, and the complex linear superfield compensator $\Upsilon$.
The relevant material can be found in ref. \cite{book}.
The action is simply ( we set $\kappa =1$ in eq. (\ref{action}) )
\beq
S_G =-\int d^4x d^4 \th~ E^{-1} = -\int d^4x d^4 \th ~ \hat{E}^{-1}
(\tilde{\Upsilon} \Upsilon)^{-1}
\label{SG}
\eeq
where $\Upsilon$ satisfies the linearity condition
$\Db^2 \Upsilon=0$ and, in "chiral" representation,
$\tilde{\Upsilon} =e^{-H}\bar{\Upsilon} e^H$, with  $H\equiv iH^a\pa_a$.
We write $\Upsilon=1+\t$ and solve the linearity condition in
terms of a spinor superfield
$\bar{\t}_\ad$
\beq
\Upsilon = 1+\t= 1+\Db^\ad \bar{\t}_\ad
\label{upsilon}
\eeq
Also,
\beq
\hat{E}^{-1}~=~ [det(1+\D)]^{-1}
=~e^{-tr~ln(1+\D)}
\label{determinant}
\eeq
where
\bea
&&\D_a^{~b}=-i\Db_\ad \D_\a^{~b} \nonumber\\
&&\D_\a^{~b} i\pa_b= e^{-H} D_\a e^H -D_\a
\label{nabla}
\eea
The action in (\ref{SG}) is invariant under local
supersymmetry transformations. The complete
gauge variation of the vector superfield $H_a$ can be written
as \cite{book}
\beq
e^{H'}=e^{i\bar{\L}}e^He^{-i\L}
\label{complete}
\eeq
where $\L\equiv \L^AiD_A$ and
\bea
&&\L^a=-i\Db^\ad L^\a  \nonumber\\
&&\L^\a=\Db^2 L^\a~~~~~~~~~~~~~~~~~~~~~~~~~~~~~~\bar{\L}^\ad=D^2 \bar{L}^\ad
\nonumber\\
&&\L^\ad=e^{-H} D^2\bar{L}^\ad e^H
{}~~~~~~~~~~~~~~~~~\bar{\L}^\a=e^{H} \Db^2 L^\a e^{-H}
\label{explicit}
\eea
with arbitrary complex $L_\a$.
[5~{}From (\ref{complete}) the infinitesimal variation is obtained
as an infinite expansion of the form
\beq
\d H= i(\bar{\L}-\L) +\frac12[i(\bar{\L}+\L),H]+
\frac{1}{12}[~[i(\bar{\L}-\L),H],H]+\dots
\label{infH}
\eeq
In the same way the infinitesimal variation of the linear
superfield $\Upsilon$ is given by \cite{book}
\beq
\d \Upsilon=(\Db_\ad \L^\ad) \Upsilon+[i\L,\Upsilon]
\label{infY}
\eeq
Using the explicit expressions in (\ref{explicit}), from
(\ref{infH}) and (\ref{infY}) we obtain respectively
\bea
&&\d H_a=D_\a \bar{L}_\ad -\Db_\ad L_\a +\frac{i}{2}(D^\b \bar{L}^\bd
+\Db^\bd L^\b) \pa_b H_a -\frac{i}{2} H^b \pa_b (D_\a \bar{L}_\ad+\Db_\ad L_\a)
\nonumber\\
&&~~~~~~~~~~~~~
-\Db^2 L^\b D_\b H_a -D^2\bar{L}^\bd \Db_\bd H_a + \dots
\label{variationH}
\eea
and
\bea
\d \t_\a&=&-\Db^2 L_\a(1+D^\b \t_\b)-\bar{L}^\ad \Db_\ad D^2 \t_\a
-\frac12 D_\a(\bar{L}^\ad \Db_\ad D^\b \t_\b) \nonumber\\
&~&-H^b i\pa_b \Db^2 L_\a +\dots
\label{variationsigma}
\eea

Having solved the linearity constraint
on $\Upsilon$ in terms of $\t_\a$, we have introduced an additional
gauge freedom under
\beq
\d  \t_\a = D^\b \L_{(\a \b )}
\label {linear}
\eeq
with arbitrary complex  $ \L_{\a \b }$. Both invariances
in (\ref{variationH}, \ref{variationsigma}) and (\ref{linear}) require
gauge fixing.

Substituting in (\ref{determinant}) the explicit expressions
in (\ref{nabla}), to quadratic order in $H_a$ we have
\bea
 \hat{E}^{-1}& =& 1-\Db_\ad D_\a H^a -\frac{i}{2}
\Db_\ad (D_\a H^c \pa_c H^a-H^c \pa_c D_\a H^a) \nonumber\\
&&~~~~~~~~~~~+\frac12 \Db_\ad
D_\a H^b \Db_\bd D_\b H^a +\frac12 (\Db_\ad D_\a H^a)^2
\eea
Thus, to this order in the supergravity fields, the action (\ref{SG}) becomes
\bea
S_G^{(2)} &=& - \int d^4x d^4 \th \left[ \frac12 H^{\a \ad}\Box H_{\a \ad}
- \frac12
 H^{\a \ad} \{ D^2, \Db^2 \}H_{\a \ad}\right.  \\
 &~&\left. + (D^\a \t_\a ) (\Db^\ad \bar{\t}_\ad )
+ (D^\a \t_\a )^2
+ (\Db^\ad \bar{\t}_\ad )^2
- H^{\a \ad}(\Db_\ad D^2 \t_\a -
D_\a \Db^2 \bar{\t}_\ad ) \right]\nonumber
\label{quadratic}
\eea
In order to fix the gauge invariances and obtain standard kinetic
terms we  first introduce instead of $\s_{\a}$ the field
\beq
\t_\a=\s_\a+z~\Db^\ad H_{\a\ad}
\label{fieldred}
\eeq
and then  introduce  two gauge--fixing functions,
corresponding to the invariances in (\ref{variationH},
\ref{variationsigma}) and (\ref{linear})
\bea
F_\a &=& \Db^\ad H_{\a \ad} +a~ \s_\a \nonumber\\
F_{\a \b} &=& D_{(\a}\s_{\b )} +b~ i \pa_{(\a}^{~~\bd}H_{\b )\bd}
+c~ D_{(\a} \Db^{~\bd}H_{\b )\bd}
\label {GFfunctions}
\eea
with suitable constants $a$, $b$, $c$,  and suitable gauge-fixing
terms. We want to achieve the double goal of cancelling cross-terms
between  $H_{\a \ad}$  and ${\s}_{\a}$,  and putting the quadratic
kinetic terms in a form which leads to simple propagators.

We start with  a  quadratic combination of these gauge-fixing functions
corresponding to a general 't Hooft gauge averaging,
\bea
L_{GF} &=&p~\bar{F}_\ad D^\a \Db^\ad F_\a+q~F_\a D^\a
\Db^\ad \bar{F}_\ad +[~r~F^\a D^2F_\a\nonumber\\
&&+\frac12 s~F^{\a \b}F_{\a \b} +\frac12 t~D^{(\b}F^{\a )}F_{\a \b}
+ {\rm h.c.}]
\label{GFgravity}
\eea
and work out all linearly independent quadratic terms in
$H^{\a \ad}$, $\s_\a$ and
$\bar{\s}_\ad$, to be added to the classical lagrangian.
In the Appendix we have listed some identities that we have used in
order to obtain a minimal set of independent quadratic quantities.
We find that with the following choice of constants,
\bea
&&s=r=0~~~~~~~~~~~~~~~~~~~~~~c=-\frac56 \nonumber\\
&&p=b=1 ~~~~~~~~~~~~~~~~~~~~~~q=\frac54 \nonumber\\
&&t=z=-\frac12 ~~~~~~~~~~~~~~~~~~~a=-\frac23
\eea
the quadratic supergravity action becomes
\beq
S_G^{(2)}+S_{GF} = - \int d^4x d^4 \th \left[
\frac12 H^{\a \ad}\Box H_{\a \ad}
+\frac49 \bar{\s}_\ad( \Db^\ad D^\a-D^\a \Db^\ad)\s_\a \right]
\eeq

Although the gauge has been fixed, and the kinetic operators are
invertible, the one
for the $\s_\a$ field will not lead to a convenient propagator.  It can
be recast in standard
form by a (non--local) field redefinition, or, equivalently, by the use of
catalyst fields
 \cite{book,GZ}.  We perform the shift
\beq
\s_\a \rightarrow \s_\a +\Db^2 D_\a \psi+ D_\a \chi~~~~~~~~~~~~~~~~~~~~~
\Db_\ad \chi=0
\label{catalyst}
\eeq
where $\psi$ is a general scalar and $\chi$ a chiral scalar superfield.
This leads to new gauge invariances
\bea
&&\d \s_\a=\Db^2 D_\a K+D_\a \L~~~~~~~~~~~~~~~~~~~~~~~~~~~(\Db_\ad\L=0)
\nonumber\\
&&\d\psi=-K~~~~~~~~~~~~~~~~~~~~~~\d\chi=-\L
\eea
Correspondingly we choose as gauge--fixing functions
\bea
&&F_1=D^\a \s_\a +a~ D^\a \Db^2 D_\a \psi+b~ D^2 \chi
+c~\Db^2 D^2\psi\nonumber\\
&&F_2=~\Db^2 \psi
\eea
and we introduce the gauge--fixing lagrangian
\beq
h~ \bar{F}_1F_1+k~\bar{F}_2 \Box F_2
\label{addGFgravity}
\eeq
With constants $a=1/2$, $b=3/2$, $h=8/9$, $k=-2/9$
the total quadratic lagrangian becomes
\beq
- \frac12 H^a \Box H_a +\frac49 \bar{\s}_\ad i\pa^{\a\ad} \s_\a-\frac23
 \bar{\chi}\Box\chi
+\frac29 \bar{\psi}\Box D^\a\Db^2 D_\a\psi
+\frac89c^2\bar{\psi}\Box \Db^2 D^2\psi
-\frac29\bar{\psi}\Box D^2 \Db^2\psi
\eeq
A final shift, with a new catalyst field, is needed in order to obtain a good
kinetic term for $\psi$:
\beq
\psi\rightarrow \psi+\r~~~~~~~~~~~~~~~~~~~~~\Db_\ad\r=0
\label{last}
\eeq
The new invariance under a chiral, opposite shift of $\psi$ and
$\r$, is fixed simply by a gauge--fixing function
$F=D^2(\psi+u\r)$ and a gauge--fixing term $w\bar{F}\Box F$.
Appropriate choices of the constants lead to a cancellation of cross
terms and to kinetic terms
\beq
-\frac29 \bar{\psi}\Box^2\psi   +\frac29 u~\bar{\r}\Box^2\r
\eeq

We note that the shifts in (\ref{fieldred}) and (\ref{catalyst}),
but not the one in (\ref{last}),
introduce additional couplings in the action (\ref{action}) between
$\psi$ and $\chi$
and the physical fields.

Although the gauge fixing in (\ref{GFgravity}, \ref{addGFgravity})
solves the problem of
providing good kinetic terms for $H^a$ and $\s^\a$, we should
caution the reader that the quantization of $\s^\a$ is more subtle
\cite{toine}.
The gauge transformation in (\ref{linear}) has zero modes,
$\L_{(\a\b)}=D^\g \L_{(\a\b\g)}$ and eventually one generates an infinite tower
of ghosts. However since these ghosts do not couple to the physical fields
they play no role.
 The propagators for the supergravity fields are then
\bea
\langle H^{\a \ad}(x, \th )H_{\b \bd}(x', \th ')  \rangle &=&
\frac{ \d^\a_{~\b} \d^\ad_{~\bd}}{\Box}
 \d^4(x-x') \d^4(\th - \th ') \nonumber\\
\langle \s_\a (x, \th ) \bar{\s} _\ad (x', \th ')  \rangle
&=& \frac94~i \frac{\pa_{\a \ad}}{\Box}
 \d^4(x-x') \d^4(\th - \th ')\nonumber\\
\langle \chi (x, \th ) \bar{\chi} (x', \th ')  \rangle
&=& \frac32\frac{1}{\Box^2}
 \d^4(x-x') \d^4(\th - \th ') \nonumber\\
\langle \psi (x, \th ) \bar{\psi} (x', \th ')  \rangle
&=& \frac92\frac{1}{\Box^2}
 \d^4(x-x') \d^4(\th - \th ')
\eea

Whereas the ghost fields introduced so far
have been effectively ignored having no interaction with the
quantum fields, this is no longer the case for the Faddeev--Popov
ghosts of the supergravity multiplet. They interact with $H_a$ and
$\s_\a$ because the gauge variations are nonlinear.
The relevant gauge variations are in (\ref{variationH}) and in
(\ref{variationsigma}). From the latter, using also (\ref{fieldred})
with $z=-\frac12$, one has
\beq
\d\s_\a=\d\t_\a+\frac12\Db^\ad \d H_{\a \ad}
\eeq
Given the field transformations, from the variation of the
gauge--fixing functions in (\ref{GFfunctions}), we construct
the Faddeev--Popov lagrangian
\beq
L_{FP}= L'^\a\d F_\a +\L'^{\a\b} \d F_{\a\b}+~h.c.
\eeq
where the variations are with respect to the
Faddeev--Popov ghosts $L_\a$, $\L_{\a\b}$  and
$L'_\a$, $\L'_{\a\b}$ are the corresponding antighosts.
By direct inspection of the quadratic and interaction terms
for the $\L_{\a\b}$, $\L'_{\a\b}$ fields, it is possible to establish
that they do not play any quantum role. Indeed, the kinetic matrix
has a triangular structure  and since  the $\L_{\a\b}$
do not interact, one cannot close a loop whenever
the $\L_{\a\b}$'s are present. Therefore it is sufficient to
consider  the part of the Faddeev--Popov lagrangian
in which only the $L_\a$ and $L'_\a$ fields appear. With a trivial
rescaling of the antighosts one obtains
\bea
L_{FP}&=& -\bar{L}'^\ad D^2 \bar{L}_\ad- \bar{L}'^\ad D_\a
\Db_\ad L^\a - L'^\a \Db^2 L_\a -L'^\a \Db_\ad D_\a \bar{L}^\ad
\nonumber\\
&~&+(D^\a\bar{L}'^\ad -\Db^\ad L'^\a)
\left[~\frac12 H^bi\pa_b (D_\a \bar{L}_\ad+\Db_\ad L_\a)
-\frac{i}{2} \pa_b H_a(D^\b\bar{L}^\bd+\Db^\bd L^\b)
\right. \nonumber\\
&~&\left.
+\Db^2 L^\b D_\b H_a +D^2 \bar{L}^\bd \Db_\bd H_a \right]
+\left[ L'^\a \left(
\Db^2 L_\a D^\b \s_\b +\bar{L}^\ad \Db_\ad D^2 \s_\a \right.
\right. \nonumber\\
&~&\left. \left.  +\frac12 D_\a(\bar{L}^\ad \Db_\ad D^\b \s_\b)
+H^b i\pa_b \Db^2 L_\a  \right) -\frac12
 L'^\a \left(
\Db^2 L_\a D^\b \Db^\bd H_b \right.\right. \nonumber\\
&~&\left. \left.+\bar{L}^\ad \Db_\ad D^2 \Db^\bd H_{\a\bd}
+\frac12 D_\a(\bar{L}^\ad \Db_\ad D^\b \Db^\bd H_b) \right)
+h.c.\right] \nonumber\\
&~&+~higher~ order~terms
\label{Faddeev}
\eea
This action has the linearized gauge invariance $\d L_\a=\L_\a$ with
$\Db_\ad \L_\a=0$ and will introduce ghosts--for--ghosts.
In order to obtain a standard quadratic
gauge--fixed kinetic term, it is convenient to already introduce
catalysts \cite{book}
as we did earlier, with the shifts
\beq
L_\a \rightarrow L_\a +D_\a U \qquad  \qquad \qquad
L'_\a \rightarrow L'_\a +D_\a U'
\eeq
The enlarged gauge invariance of the resulting action is then
\bea
&&\d L_\a=\L_\a +D_\a L \qquad \qquad \qquad \Db_\ad \L_\a=0
\nonumber\\
&&\d U=-L
\eea
We choose as gauge--fixing functions
\bea
&&F=D^\a L_\a +D^2U +\Db^2 (dU+e\bar{U})\nonumber\\
&&F_\a =\Db^2[L_\a+D_\a(U+ \bar{U})]
\eea
and similar terms $F'$, $F'_\a$, and add to (\ref{Faddeev})
the gauge--fixing term
\beq
-\bar{F}' F-F'\bar{F}
+~\bar{F}'^\ad \frac{\Db^2}{\Box} \bar{F}_\ad
+ ~F'^\a \frac{D^2}{\Box} F_\a
\eeq
In this way the quadratic part of the lagrangian becomes
\bea
L^{(2)}_{FP}&= &-\bar{L}'^\ad i\pa_{\a\ad} L^\a -L'^\a
i\pa_{\a\ad}\bar{L}'^\ad+\left[- \bar{U}'D^\a \Db^2D_\a \bar{U}
-\bar{U}'D^\a \Db^2D_\a U
\right. \nonumber\\
&~&
 -d'e~ \bar{U}'D^2\Db^2\bar{U}
-e'd~\bar{U}'\Db^2 D^2\bar{U}-dd'~\bar{U}'D^2\Db^2U \nonumber\\
&~&\left.+(1-ee')\bar{U}'\Db^2D^2U+~h.c.\right]
\eea
Again the shifts above introduce couplings between the $U$ fields  and $H^a$.
The U lagrangian is not yet in standard form, but this can be easily
achieved introducing some chiral catalyst fields. We refer the interested
reader to Appendix B of ref. \cite{GZ}, where the procedure is
spelled out in detail. We emphasize that the shifts in (5.37) and
the subsequent shifts
in the catalysts $U$, do lead to unavoidable couplings of these new
fields to the supergravity field.

The quantization of the system is now complete.

\sect{Conclusions}

The motivation for undertaking this work was to provide
all the ingredients necessary for quantitative, perturbative
calculations in $N=2$ matter--supergravity systems. Having reached this goal,
the next step would be to perform an actual calculation.
We are studying the one--loop four--point amplitude in order to verify the
argument of section 2 by establishing that the quadratic divergences cancel.
This requires the computation of
one--loop irreducible contributions
and also self--energies and vertex corrections to be inserted in tree
diagrams using all the interaction terms to first and second
order in the supergravity fields, $H_a$
and $\s_\a$. These terms are contained  not only in the superdeterminant
$E^{-1}$, but also in
the covariant derivatives, torsions, etc.
Even if conceptually
straightforward, in practice this calculation is far too complicated
to be performed in a naive, brute force way. In order to maintain the
number of contributions under control, one has to use some ingenuity,
and it is indeed possible to streamline the procedure somewhat.

 We hope to report on this calculation in a not
too distant future \cite{GSZ}.

\vspace{2cm}
{\bf Acknowledgments} We thank the Physics Department of Harvard
University for hospitality.
This research was supported in part by NSF Grant PHY--92--22318,
by the European Commission TMR program ERBFMRX--CT96--0045, in which
A. S. and D. Z. are associated to Torino, and by INFN.

\appendix
\section{ Appendix }
\setcounter{equation}{0}
We collect in this Appendix some identities that allow one to identify
a minimal set of independent expressions quadratic in the supergravity
field $H_a$.
Besides the terms $H^a\Box H_a$, $H^aD^2\Db^2H_a$, $H^a D^\b\Db^2 D_\b H_a$
we have included in the minimal set the following expressions:
$H^a\pa_a\pa_b H^b$, $H^a\pa_a D^\b\Db^\bd H_b$ and
$H_{\a\bd} D^\a \Db^2 D_\b H^b$. In terms of these quantities
we have obtained
\bea
&&H_a \pa^{\a\bd} \pa^{\b\ad} H_b=H_a \pa^a\pa^b H_b -H^a\Box H_a
\nonumber\\
&&H_a \pa^a\Db^\bd D^\b H_b=H_a\pa^ai\pa^b H_b -
H_a\pa^a D^\b\Db^\bd H_b\nonumber\\
&&H_a\pa^{\a\bd} D^\b\Db^\ad H_b=H_a\pa^ai\pa^b H_b
-H_a\pa^aD^\b\Db^\bd H_b \nonumber\\
&&~~~~~~~~~~~~~~~~-iH^a\Box H_a+iH^a D^2\Db^2H_a
-iH_{\b\ad}\Db^\ad D^2\Db_\bd H^b\nonumber\\
&&H_a\pa^{\b\ad}\Db^\bd D^\a H_b=H_a\pa^a D^\b\Db^\bd H_b
-iH^a \Box H_a+iH^a D^2\Db^2 H_a -
iH_{\a\bd}D^\a\Db^2 D_\b H^b\nonumber\\
&& H_{\a\bd}D_\b\Db^2 D^\a H^b=H_{\a\bd}D^\a\Db^2 D_\b H^b-
H^a D^\b\Db^2 D_\b H_a
\eea
For the gauge fixing in (\ref{GFgravity}) we obtain
\beq
\bar{F}_\ad D^\a \Db^\ad F_\a=H^aD^2 \Db^2 H_a+a^2~
\bar{\s}_\ad D^\a \Db^\ad \s_\a +a~\bar{\s}^\ad D^\a \Db^2 H_a
-a~\s^\a \Db^\ad D^2 H_a
\eeq

\bea
F_\a D^\a \Db^\ad \bar{F}_\ad &=& H_ai\pa^a D^\b \Db^\bd H_b
-H_{\b\ad} \Db^\ad D^2 \Db_\bd H^b
+a^2~ \s_\a D^\a \Db^\ad \bar{\s}_\ad
-a~\s_\a D^\a i\pa^b H_b
\nonumber\\
&~&
+a~\s^\a D^2 \Db^\ad H_a +a~ \bar{\s}_\ad \Db^\ad i\pa^b H_b
-a~\bar{\s}^\ad \Db^2D^\a H_a
\eea

\beq
F^\a D^2 F_\a= -H_a \Db^\ad D^2 \Db_\bd H^{\a\bd} +
2a~ \s^\a D^2 \Db^\ad H_a +a^2~ \s^\a D^2 \s_\a~~~~~~~~~~~~~~~~~~~
{}~~~~~~~~
\eeq

\bea
\frac12 F^{\a\b} F_{\a\b}&=& 3\s^\a D^2 \s_\a +2b~ \s^\a D_\a i\pa_b H^b
-4b~ \s^\b \Db^\bd D^2 H_b +(4b+6c) \s^\b D^2 \Db^\bd H_b
\nonumber\\
&~&+2b^2~ H^a \Box H_a -(b^2+2bc) H_a \pa^a \pa^b H_b
-(3c^2+2cb)H_b \Db^\bd D^2 \Db_\ad H^{\b\ad}
\nonumber\\
&~&+4bc~ H^a D^2 \Db^2 H_a
-2bc~H_a i\pa^a D^\b \Db^\bd H_b -2bc~ H_{\a\bd} D^\a \Db^2 D_\b H^b
\eea

\bea
\frac12 D^\b F^\a F_{\a\b}&=& 3a~\s^\a D^2\s_\a +ab~ \s^\a D_\a i\pa_b H^b
+(3+2ab+3ac)\s^\a D^2 \Db^\ad H_a \nonumber\\
&~&-2ab~ \s^\a \Db^\ad D^2 H_a +2b~ H^a D^2 \Db^2 H_a
-(2b+3c) H_b \Db^\bd D^2 \Db_\ad H^{\b\ad}\nonumber\\
&~&+b~ H_a i\pa^a D^\b \Db^\bd H_b
\eea


\end{document}